\documentclass[showpacs,preprintnumbers,amsmath,amssymb,aps, prd]{revtex4}
\usepackage{graphicx}
\usepackage{grffile}
\usepackage{mathrsfs, mathtools}
\usepackage{caption}
\usepackage{float}
\usepackage{xcolor}
\usepackage{changepage}
\usepackage{dcolumn}
\usepackage{bm}
\usepackage{graphicx,subfigure,epsfig}
\usepackage{multirow}

\def\mn{_{\mu\nu}}
\def\MN{^{\mu\nu}}

\def\a{\alpha}
\def\b{\beta}
\def\f{\frac}

\def\l{\mathcal{L}}

\def\ab{\overline{a}}

\def\bb{\overline{b}}
\def\tb{\tilde{\beta}}
\def\c{\cite}
\def\r{\ref}
\def\ar{A(r)}
\def\br{B(r)}
\def\fx{F(x)}
\def\hx{H(x)}
\def\s{Schwarzschild }
\newcommand\be{\begin{equation}}
\newcommand\ee{\end{equation}}
\newcommand\ba{\begin{eqnarray}}
\newcommand\ea{\end{eqnarray}}
\newcommand\bt{\bibitem}
\newcommand\nn{\nonumber}
\newcommand\lt{\left}
\newcommand\rt{\right}
\newcommand\pt{\partial}
\newcommand\tx{\text}

\begin{document}
\title{Black hole surrounded by perfect fluid dark matter with a background Kalb-Ramond field}
\author{Sohan Kumar Jha}
\email{sohan00slg@gmail.com}
\affiliation{Department of Physics, Chandernagore College, Chandernagore, Hooghly, West
Bengal, India}

\date{\today}
\begin{abstract}
\begin{center}
Abstract
\end{center}
With an intent to explore the interplay between the Lorentz symmetry breaking (LSB) and the presence of dark matter (DM), we obtain a static and spherically symmetric black hole (BH) solution in the background of nonminimally coupled Kalb-Ramond (KR) field surrounded by perfect fluid dark matter (PFDM). The KR field is frozen to a non-zero vacuum expectation value (VEV) that breaks the particle Lorentz symmetry spontaneously. We explore scalar invariants, Ricci Scalar, Ricci squared, and Kretschmann Scalar, to probe the nature of singularities in the obtained solution. We then study strong gravitational lensing in the background of our BH, i.e., KRPFDM BH, revealing the adverse impact of LSB parameter $\a$ and PFDM parameter $\b$ on the lensing coefficients. The significant effect of our model parameters is evident in strong lensing observables. Bounds on the deviation from \s, $\delta$, for supermassive BHs (SMBHs) $M87^*$ and $SgrA^*$ from the EHT, Keck, and VLTI observatories are then utilized to put our BH model to the test and extract possible values of model parameters $\a$ and $\b$  that generate theoretical predictions in line with experimental observations within $1\sigma$ confidence level. Our study sheds light on the combined effect of LSB and PFDM and may be helpful in finding their signature.\\      

\end{abstract}
\maketitle
\section{Introduction}
Lorentz symmetry forms the bedrock of general relativity (GR) and standard model and has widely been ascribed as the central pillar of modern physics. Despite the symmetry being confirmed by numerous observations, numerous theories motivated by non-commutative field theory, string theory, and others predict that the symmetry is broken at the fundamental scale [\citenum{Kostelecky1989a}-\citenum{Cohen2006}]. We may have either spontaneous or explicit violation of Lorentz symmetry. In the case of explicitly broken, the invariance of Lagrange density under Lorentz transformation breaks down, and physical laws no longer remain the same in all inertial frames. On the other hand, Lorentz symmetry being broken spontaneously does not affect the invariance of Lagrange density but breaks the Lorentz symmetry in the ground state. The standard model extension (SME) in this context has been developed to incorporate LSB \cite{Kostelecky2004a}. Within this extension, one way to incorporate LSB is through the bumblebee model where a vector field called bumblebee field, non-minimally coupled to gravity, attains a non-zero VEV \cite{Kostelecky1989a, Kostelecky1989, Kostelecky1989b, Bailey2006, Bluhm2008a}. In article \c{bm}, authors have obtained a Schwarzschild-like solution. Please see [\citenum{Ovgun2018}-\citenum{Amarilo2023}] for some of the notable works related to BHs in this model. We may also induce LSB by considering a rank-two antisymmetric KR field, non-minimally coupled to gravity, that acquires a non-zero VEV \c{action1}. Please see [\citenum{Kalb1974}-\citenum{Chakraborty2016}] for more details regarding the KR field. An exact solution in this model was obtained in \cite{action2}. Please refer [\citenum{Atamurotov2022}-\citenum{Maluf2022a}] for various studies in this regard. A different solution in this model was also obtained in \c{kr}.\\
The existence of DM is an enigma that has been intriguing researchers for some time now. It is expected that matter fields surround astrophysical BHs, and DM is considered a possible candidate. In the quest for the existence of DM, Observations vis-a-vis elliptical and spiral galaxies provided the first breakthrough \c{rubin}. One study pegs DM's contribution to a galaxy's mass to around $90\%$ \c{persic}. Numerous evidence posits that DM halos embed astrophysical BHs \cite{akiyamal1,akiyamal6}. It thus becomes imperative to reckon the contribution of DM near the galactic center \c{sofue, boshkaye}. One can find numerous DM profiles that can help incorporate DM effect [\citenum{kiselev}-\citenum{rayimbaev}]. Here, we have considered the PFDM model, where the dark matter is assumed to be a perfect fluid. This model can explain rotation curves with regard to spiral galaxies \c{pf1}. Please see \c{pf2, pf3, pf4, pf5} for some recent works with PFDM. \\
Gravitational lensing in the strong field limit has been the subject of intense research as the relativistic images encode imprints of intricate properties of the region just outside the event horizon. It is possible to extract valuable information regarding strong fields from these gravitationally lensed relativistic images. As modified theories of gravity must be consistent with GR while taking weak field limits, it is imperative to probe strong fields to gauge deviations from GR. It is in this regard that gravitational lensing in the strong-field limit proves to be a potent tool. Electromagnetic radiation, as it passes by a compact astrophysical object, experiences deflection, commonly known as gravitational lensing, and the compact object is referred to as a gravitational lens. The seminal work \c{dar} by Darwin, where the author for the first time applied the concept of gravitational lensing to the \s BH, became the foundation for further developments in the future. Virbhadra and Ellis, in their pioneering work \c{vir}, derived the gravitational lens equation. With this as a foundation, Bozza and others in \c{BOZZA} have further developed the analytical method and applied it to a wider range of spacetimes. Please see [\citenum{BOZZA1}-\citenum{lens13}] for more details on the application of the method to various spacetimes. Here, we apply the method to find the signature of DM and LSB. Please refer \c{id1, id2, id3} for the application of gravitational lensing to find the signature of DM and see \c{ikr1} where the effect of LSB for the spacetime \c{kr} is studied. Experimental observations provide an excellent avenue for extracting valuable information and testing our model. We will utilize bounds on the deviation from \s BH, $\delta$, for the SMBHs $M87^*$ \c{M871, M872} and $Sgr A^*$ \c{keck, vlti1, vlti2} to gauge the feasibility of our model and extract bounds on free parameters. \\
In this article, Section II presents the derivation of the static and spherically symmetric metric. Sections III and IV are devoted to studying strong gravitational lensing and related observables, respectively. In Section V, we constrain our free parameters utilizing experimental observations, and with concluding remarks in Section VI, we end our article. We have used $G=c=M=1$ in this article.
\section{BH surrounded by PFDM with a background KR field}
LSB considered in this article occurs due to non-zero VEV of the KR field $B\mn$, which is an anti-symmetric tensor of rank two. The KR field is non-minimally coupled to gravity. We intend to find the black hole solution with a background KR field surrounded by PFDM (KRPFDM) whose action reads \c{action1, action2}
\begin{widetext}
\begin{align}\label{action}
S=\int d^4x\sqrt{-g}\bigg[\frac{1}{2\kappa}\bigg(R-2\Lambda+\varepsilon\, B^{\mu\lambda}B^\nu\, _\lambda R_{\mu\nu}\bigg)-\frac{1}{12}H_{\lambda\mu\nu}H^{\lambda\mu\nu}-V(B_{\mu\nu}B^{\mu\nu}\pm b^2)+\mathcal{L}_{dm}\bigg],
\end{align}
\end{widetext}
where $\kappa=8\pi G$, $G$ being the Newtonian gravitational constant, $\varepsilon$ is the coupling constant between gravitation and the KR field, $b^2$ is a real positive constant, $\Lambda$ is the cosmological constant, and the KR field strength is given by $H_{\mu\nu\rho}\equiv \pt_{[\mu}B_{\nu\rho]}$. In Eq. (\r{action}), $\mathcal{L}_{dm}$ is the Lagrangian density for PFDM. The self-interacting potential $V(B_{\mu\nu}B^{\mu\nu}\pm b^2)$ ensures spontaneous breaking of Lorentz symmetry and generates a non-zero VEV of the KR field $\langle B_{\mu\nu} \rangle=b_{\mu\nu}$ such that the constant norm condition $b_{\mu\nu}b^{\mu\nu}=\mp b^2$ is satisfied. This leads to the null field strength of the KR field.\\
Varying the action (\r{action}) with respect to $g^{\mu \nu}$ leads to the following field equations:
\ba
R_{\mu \nu }-\frac{1}{2}g_{\mu \nu }R+\Lambda  g_{\mu \nu }=\kappa \lt( T^{\tx{KR}}_{\mu\nu}+T^{\tx{M}}_{\mu\nu}\rt),
\label{fe}
\ea
where $T^{\tx{DM}}_{\mu\nu}$ is the energy-momentum tensor of PFDM given by
\ba\nn
T_{\mu}^{\nu(\tx{DM})}&=&diag\lt[-\rho,p,p,p\rt]\\\nn
&=&diag\lt[\f{\b}{8\pi r^3}, \f{\b}{8\pi r^3}, -\f{\b}{16\pi r^3}, -\f{\b}{16\pi r^3} \rt],
\label{tdm}
\ea
and
\ba
\kappa T^{\tx{KR}}_{\mu\nu} \!&\!=\!&\!\frac{1}{2} H_{\mu \alpha \beta } H_{\nu }{}^{\alpha \beta } \!-\! \frac{1}{12} g_{\mu \nu } H^{\alpha \beta \rho } H_{\alpha \beta \rho }\!+\!2V'(X) B_{\alpha\mu}B^{\alpha}{}_\nu  \nn\\
\!&\!-\!&\! g_{\mu\nu}V(X)+ \varepsilon \bigg[\frac{1}{2} g_{\mu \nu } B^{\alpha \gamma } B^{\beta }{}_{\gamma }R_{\alpha \beta } - B^{\alpha }{}_{\mu } B^{\beta }{}_{\nu }R_{\alpha \beta }\nn\\
\!&\!-\!&\!  B^{\alpha \beta } B_{\nu \beta } R_{\mu \alpha }-B^{\alpha \beta } B_{\mu \beta } R_{\nu \alpha }+\frac{1}{2} \nabla _{\alpha }\nabla _{\mu }\left(B^{\alpha \beta } B_{\nu \beta }\right)
\nn\\
\!&\!+\!&\!
\frac{1}{2} \nabla _{\alpha }\nabla _{\nu }\left(B^{\alpha \beta } B_{\mu \beta }\right)-\frac{1}{2}\nabla ^{\alpha }\nabla _{\alpha }\left(B_{\mu }{}^{\gamma }B_{\nu \gamma } \right)
\nn\\
\!&\!-\!&\! \frac{1}{2} g_{\mu \nu } \nabla _{\alpha }\nabla _{\beta }\left(B^{\alpha \gamma } B^{\beta }{}_{\gamma }\right)\bigg].
\label{tkr}
\ea
where the prime denotes the derivative of the corresponding functions with respect to the argument. The Bianchi identities ensure conservation of the combined tensor $T^{\tx{KR}}_{\mu\nu}+T^{\tx{M}}_{\mu\nu}$. We, in this article, consider the cosmological constant $\Lambda$ to be zero.\\
Deriving a static and spherically symmetric solution to the field equations (\r{fe}) is our intention. To this end, we consider the following ansatz:
\be
ds^2=-A(r)dt^2+B(r)dr^2+r^2 d\theta^2+r^2 \sin^2\theta d\phi^2.
\label{trial}
\ee
We consider a pseudoelectric configuration of the KR field where the only non-zero components of the field are $b_{01}$ and $b_{10}$. The constant norm condition $b_{\mu\nu}b^{\mu\nu}=\mp b^2$ yield:
\be
b_{01}=-b_{10}=|b| \sqrt{\f{\ar \br}{2}}.
\ee
The KR field is assumed to be frozen to its VEV. \\
Field equations (\r{fe}) for the ansatz (\r{trial}) result in the following equations:
\ba
&&-\frac{r B'(r)+(B(r)-1) B(r)}{r^2 B(r)^2}=\frac{\alpha  \left(r^2 B(r) A'(r)^2+r A(r) \left(r A'(r) B'(r)-2 B(r) \left(r A''(r)+A'(r)\right)\right)+2 A(r)^2 B(r)\right)}{2 r^2 A(r)^2 B(r)^2}+\f{\b}{r^3},\label{00}\\\nn
&&\frac{r A'(r)+A(r) (-B(r))+A(r)}{r^2 A(r) B(r)}=\frac{\alpha  \left(r^2 B(r) A'(r)^2+r^2 A(r) \left(A'(r) B'(r)-2 B(r) A''(r)\right)+2 A(r)^2 \left(r B'(r)+B(r)\right)\right)}{2 r^2 A(r)^2 B(r)^2}+\f{\b}{r^3},\label{11}\\
\\\nn
&&\frac{-r B(r) A'(r)^2+A(r) \left(2 B(r) \left(r A''(r)+A'(r)\right)-r A'(r) B'(r)\right)-2 A(r)^2 B'(r)}{4 r A(r)^2 B(r)^2}\\
&&=-\frac{\alpha  \left(r B(r) A'(r)^2+A(r) \left(r A'(r) B'(r)-2 B(r) \left(r A''(r)+A'(r)\right)\right)+2 A(r)^2 B'(r)\right)}{4 r A(r)^2 B(r)^2}-\f{\b}{2r^3},\label{22}
\ea
where the parameter $\a=\f{\varepsilon b^2}{2}$ embodies the LSB effect due to the non-zero VEV of the KR field. Deducting Eq. (\r{00}) from Eq. (\r{11}) yields the following relation between $\ar$ and $\br$:
\be
B(r) A'(r)+A(r) B'(r)=0,
\ee
resulting in $\br=\f{1}{\ar}$. This conjoined with Eq. (\r{00}) provides the desired metric as
\be
ds^2=-A(r)dt^2+\f{dr^2}{\ar}+r^2 d\theta^2+r^2 \sin^2\theta d\phi^2,
\label{kpf}
\ee
with $\ar=\f{1}{1-\a}-\f{2M}{r}+\f{\b}{(1-\a)r}log\f{r}{|\b|}$. The above metric in the limit $\b \rightarrow 0$ reduces to that derived in \c{kr} and an additional limit $\a \rightarrow 0$ produces the well-known \s metric. The KRPFDM BH metric presented in (\r{kpf}) has two conspicuous singularities: one at $r=0$ and another at $\ar=0$. The solution of the equation $\ar=0$ provides the position of the event horizon as:
\be
r_h=\beta  ProductLog\left(e^{-\frac{2 (\alpha -1) M}{\beta }}\right).\label{rh}
\ee
In the limit $\a,\, \b \rightarrow 0$, the above expression produces $r_h=2M$, the event horizon for a \s BH. The qualitative impact of the parameters $\a$ and $\b$ will be studied in the next section, where the effect of these parameters on lensing will be explored in detail. To probe further the nature of above-mentioned singularities, it is imperative to consider scalar invariants whose expressions are given as follows:
\ba\nn
&&\tx{Ricci Scalar}=R=\frac{\beta +2 \alpha  r}{(\alpha -1) r^3},\\\nn
&&\tx{Ricci squared}=R\mn R\MN=\frac{5 \beta ^2+4 \alpha ^2 r^2+8 \alpha  \beta  r}{2 (\alpha -1)^2 r^6},\\\nn
&&\tx{Kretschmann Scalar}=K=\\\nn
&&\frac{13 \beta ^2+48 (\alpha -1)^2 M^2-40 (\alpha -1) \beta  M+4 \beta  \log \left(\frac{r}{\beta }\right) \left(-5 \beta +12 (\alpha -1) M+2 \alpha  r+3 \beta  \log
\left(\frac{r}{\beta }\right)\right)+16 \alpha  (\alpha -1) M r+4 \alpha ^2 r^2}{(\alpha -1)^2 r^6}.\\
\label{inv}
\ea
It is evident from the above expressions that the singularity at $r=0$ is an essential one, whereas the singularity at $r=r_h$ is a removable one. The following expressions provide the non-zero components of the Ricci tensor:
\ba
R_{tt}&=&-\frac{\beta  \left(2 (\alpha -1) M+\beta  \log \left(\frac{r}{\beta }\right)+r\right)}{2 (\alpha -1)^2 r^4},\\\nn
R_{rr}&=&\frac{\beta }{2 r^2 \left(2 (\alpha -1) M+\beta  \log \left(\frac{r}{\beta }\right)+r\right)},\\\nn
R_{\theta \theta}&=&\frac{\beta +\alpha  r}{(\alpha -1) r},\\\nn
R_{\phi \phi}&=&\frac{\sin ^2(\theta ) (\beta +\alpha  r)}{(\alpha -1) r},
\ea
signifying our metric being non-Ricci flat. We now move on to explore the impact of parameters $\a$ and $\b$ on strong gravitation lensing.
\section{strong gravitational lensing by krpfdm BHs}
Strong gravitational lensing provides an excellent avenue for exploring the nature of the underlying spacetime, whose imprints can be detected through lensing observables. In this section, we study strong gravitational lensing by KRPFDM BHs using the prescription given in \cite{BOZZA, BOZZA1, BOZZA2} to evince the impact of parameters $\a$ and $\b$ on observables related to the strong gravitational lensing. To this end, we rewrite the metric (\r{kpf}), confining ourselves only to the equatorial plane, as
\begin{equation}
d\tilde{s}^{2}=(2M)^{-2}ds^{2}=-F(x) dt^{2}+F(x)^{-1} dx^{2}+H(x) d \phi^{2}, \label{kpf1}
\end{equation}
where $x=\f{r}{2M}$, $\tb=\f{\b}{2M}$, and
\be
\fx=\f{1}{1-\a}-\f{1}{x}+\f{\tb}{(1-\a)r}log\f{x}{|\tb|} \quad \tx{and} \quad \hx=x^2.
\ee
Solving the equation $\fx=0$ yields the following expression of the event horizon:
\be
x_h=\tilde{\beta } ProductLog\left(e^{\frac{1}{\tilde{\beta }}-\frac{\alpha }{\tilde{\beta }}}\right),
\ee
which can also be obtained from Eq. (\r{rh}). Fig. (\r{xh}) depicts a variation of the event horizon with $\a$ and $\tb$. $x_h$ decreases linearly with the LSB parameter $\a$. However, its variation with the PFDM parameter $\tb$ exhibits critical behavior, with the existence of a local minimum that varies with $\a$. For example, at $\a=- 0.2$ the minima occurs at $\tb=0.32273$ with $x_h=0.87727$, at $\a=0$ the minima occurs at $\tb=0.268941$ with $x_h=0.731059$, and at $\a=0.2$ the minima occurs at $\tb=0.215153$ with $x_h=0.584847$. This displays an adverse impact of increasing $\a$ on the minimum value of $x_h$ and its position.
\begin{figure}[H]
\begin{center}
\begin{tabular}{cc}
\includegraphics[width=0.4\columnwidth]{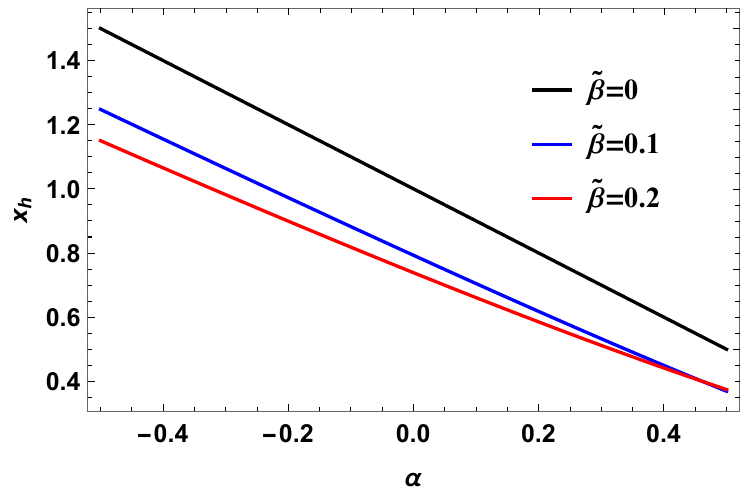}&
\includegraphics[width=0.4\columnwidth]{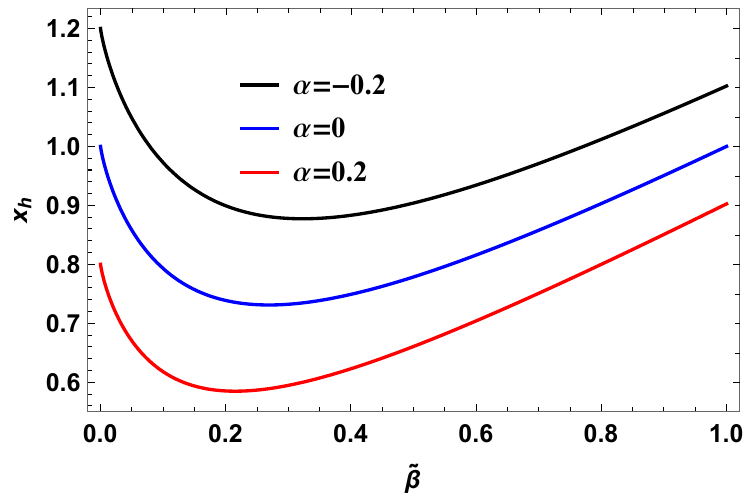}
\end{tabular}
\caption{Variation of event horizon with $\a$ for different values of $\tb$ (left panel) and with $\tb$ for different values of $\a$ (right panel). }\label{xh}
\end{center}
\end{figure}
We now move on to derive the differential equation governing null geodesics. The Lagrangian associated with the metric (\r{kpf1}) is
\be
\l=\f{1}{2}\lt(-\fx \dot{t}^2+\f{\dot{x}^2}{\fx}+\hx \dot{\phi}^2\rt),\label{l}
\ee
where $\dot{t}=\f{dt}{d\lambda}$ and $\dot{\phi}=\f{d\phi}{d\lambda}$, $\lambda$ being the affine parameter. With no dependence of the Lagrangian on $t$ and $\phi$, we have two conserved quantities associated with null geodesics: the energy $E$ and the angular momentum $L$ given by
\be
E=-\f{d\l}{d\dot{t}}=\fx \dot{t} \quad \tx{and} \quad L=\f{d\l}{d\dot{\phi}}=\hx \dot{\phi}.
\ee
This conjoined with the equation $d\tilde{s}=0$ for null geodesics provide the following equation:
\be
\dot{x}^2=E^2-\f{L^2\fx}{\hx}=E^2-V(x),
\ee
where $V(x)=\f{L^2\fx}{\hx}$ is the potential that governs the photon trajectory. For circular orbits of radius $x_m$, the potential satisfies the conditions $V(x_m)=E^2$ and $\f{dV}{dx}|_{x=x_m}=0$. The second condition provides the following equation:
\be
H'(x)\fx=F'(x)\hx,
\ee
where prime implies differentiation with respect to $x$. The above equation yields the following expression for the photon radius as
\be
x_m=\frac{3}{2} \tilde{\beta } ProductLog\left(\frac{2}{3} e^{-\frac{\alpha }{\tilde{\beta }}+\frac{1}{\tilde{\beta }}+\frac{1}{3}}\right).
\ee
In the limit $\a\rightarrow 0$ and $\tb \rightarrow 0$, the above expression produces $x_m=1.5$, which corresponds to the \s case. Its variation with $\a$ and $\tb$ are shown in Fig. (\r{xm}). Impact of $\a$ and $\tb$ on $x_m$ is similar to the event horizon case. Here, at $\a=-0.2$, the photon radius reaches its minimum value of 1.28996 at $\tb=0.340029$, $x_m$ attains a minimum value of 1.07496 at $\tb=0.283357$ when $\a=0$, and at $\a=0.2$ the radius reaches its minimum value of 0.859971 at $\tb=0.226686$.
\begin{figure}[H]
\begin{center}
\begin{tabular}{cc}
\includegraphics[width=0.4\columnwidth]{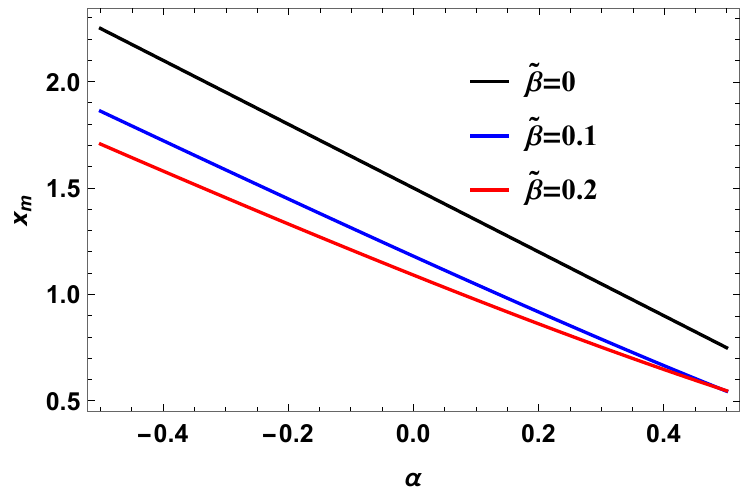}&
\includegraphics[width=0.4\columnwidth]{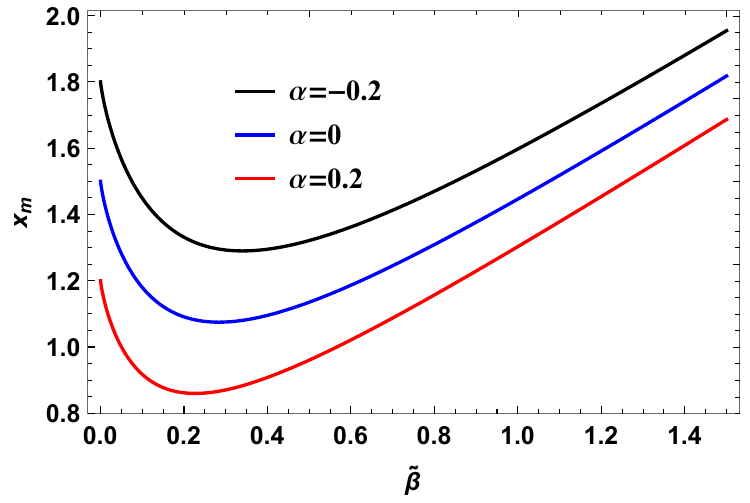}
\end{tabular}
\caption{Variation of photon radius with $\a$ for different values of $\tb$ (left panel) and with $\tb$ for different values of $\a$ (right panel). }\label{xm}
\end{center}
\end{figure}
For an impact parameter $b$, there exists a minimum distance $x_0$ that satisfies the condition:
\be
\f{dx}{d\phi}=0 \quad \Rightarrow \quad b=\f{L}{E}=\sqrt{\f{H(x_0)}{F(x_0)}}.
\ee
When $x_0=x_m$, we get the limiting value of the impact parameter $b_m$. Any photon with impact parameter $b<b_m$ will be swallowed by the BH. Fig. (\r{bm}) elucidates variation of the critical impact parameter with $\a$ and $\tb$. Its similarity with those observed in the case of the event horizon and photon radius is conspicuous.
\begin{figure}[H]
\begin{center}
\begin{tabular}{cc}
\includegraphics[width=0.4\columnwidth]{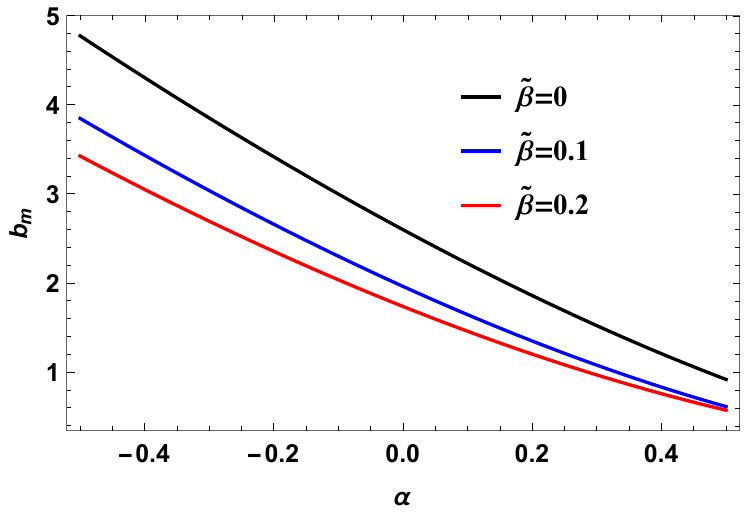}&
\includegraphics[width=0.4\columnwidth]{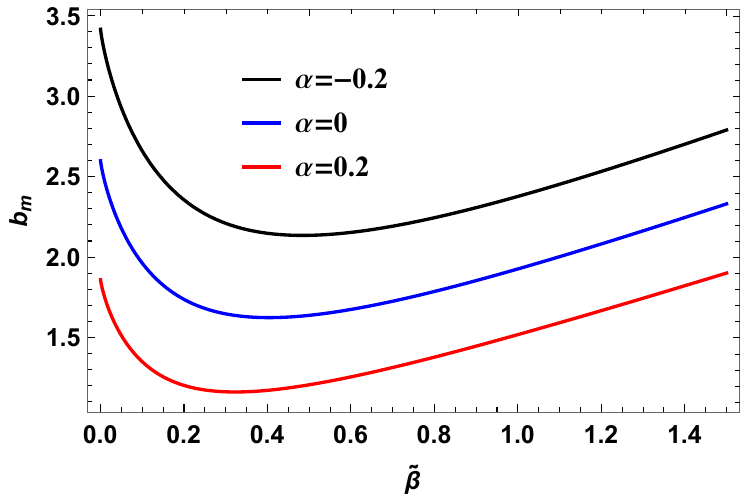}
\end{tabular}
\caption{Variation of critical impact parameter $b_m$ with $\a$ for different values of $\tb$ (left panel) and with $\tb$ for different values of $\a$ (right panel). }\label{bm}
\end{center}
\end{figure}
Following articles \c{vir, wein}, the expression for the deflection angle is
\begin{equation}
\alpha_D\left(x_{0}\right)=I\left(x_{0}\right)-\pi,\label{def}
\end{equation}
where
\begin{equation}
I\left(x_{0}\right)=\int_{x_{0}}^{\infty}\frac{2}{\sqrt{F(x)H(x)}
\sqrt{\frac{F(x_{0})H(x)}{H(x_{0})F(x)}-1}}dx.\label{io}
\end{equation}
The above expression diverges at $x_0=x_m$. Introducing the variable $z=1-\f{x_0}{x}$ and following the procedure illustrated in \c{BOZZA}, the deflection angle is given by
\begin{eqnarray}
&& \gamma_D(b)=-\overline{a} \log \left( \frac{b}{b_m} -1
\right) +\overline{b}+O(b-b_m) ,\label{alphab}\\\nn%
\text{where}\\
&& \overline{a}=\frac{a}{2}=\frac{\mathcal{R}(0,x_m)}{2\sqrt{a_2(x_m)}}, \label{afinal}\\%
&& \overline{b}=-\pi+\bar{b}_\mathcal{R}+\overline{a}\log{\frac{2H^2(x_m)a_2(x_m)}{F(x_m)x_m^4}}\label{bfinal},\\\nn
\text{with}\\\nn
&& \mathcal{R}(z,x_m)=\frac{2x^{2}\sqrt{H(x_0)}}{x_{0}H(x)},\\\nn
&&a_{2}(x_0) = \frac{1}{2}\left[\frac{\lt(2x_{0}H(x_{0})-2x_{0}^{2}H^{\prime}(x_{0})\rt)\lt(H^\prime(x_0) F(x_0)-F^\prime(x_0) H(x_0)\rt)}{H^{2}(x_{0})}+\frac{x_{0}}{H(x_{0})}\left(H^{\prime\prime}(x_{0})F(x_{0})-F^{\prime\prime}(x_{0})F(x_{0})\right)\right],\\\nn
&&g(z,x_0)=\mathcal{R}(z,x_0)f(z,x_0)-\mathcal{R}(0,x_m)f_0(z,x_0),\\\nn
&& I_\mathcal{R}(x_0)=\int\limits_0^1 g(z,x_m) dz+O(x_0-x_m) \quad \text{and} \quad \bar{b}_\mathcal{R}= I_\mathcal{R}(x_m).
\end{eqnarray}
In the absence of the KR field and PFDM, we obtain $\ab=1$ and $\bb=-0.40023$, the same values obtained in \c{BOZZA} for the \s BH. With the requisite expressions at our disposal, we now explore the qualitative aspect of the dependence lensing coefficients $\ab$ and $\bb$ display with respect to our model parameters $\a$ and $\b$. Fig. (\r{abar}) and (\r{bbar}) display an adverse impact of increasing either $\a$ or $\b$ on both the lensing coefficients.
\begin{figure}[H]
\begin{center}
\begin{tabular}{cc}
\includegraphics[width=0.4\columnwidth]{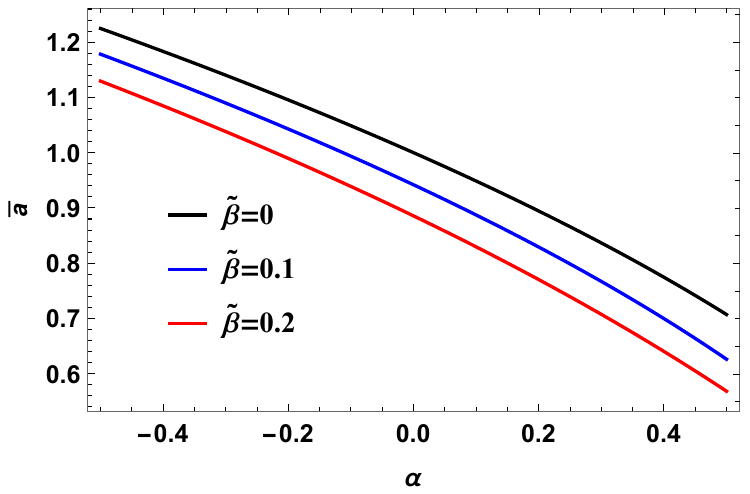}&
\includegraphics[width=0.4\columnwidth]{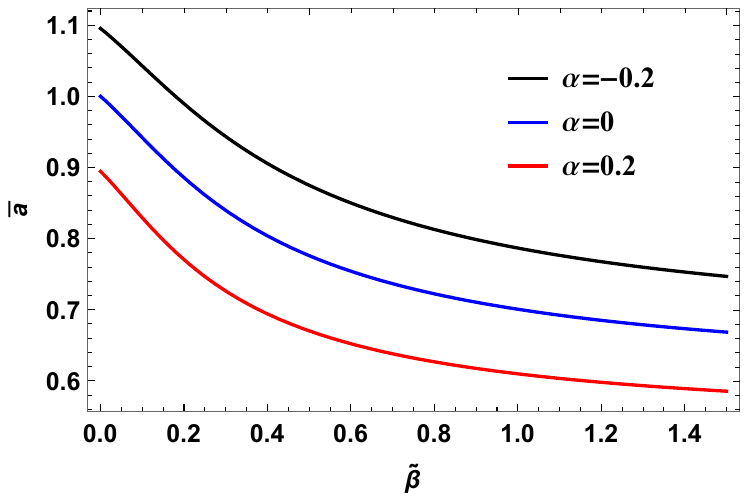}
\end{tabular}
\caption{Variation of lensing coefficient $\ab$ with $\a$ for different values of $\tb$ (left panel) and with $\tb$ for different values of $\a$ (right panel). }\label{abar}
\end{center}
\end{figure}
\begin{figure}[H]
\begin{center}
\begin{tabular}{cc}
\includegraphics[width=0.4\columnwidth]{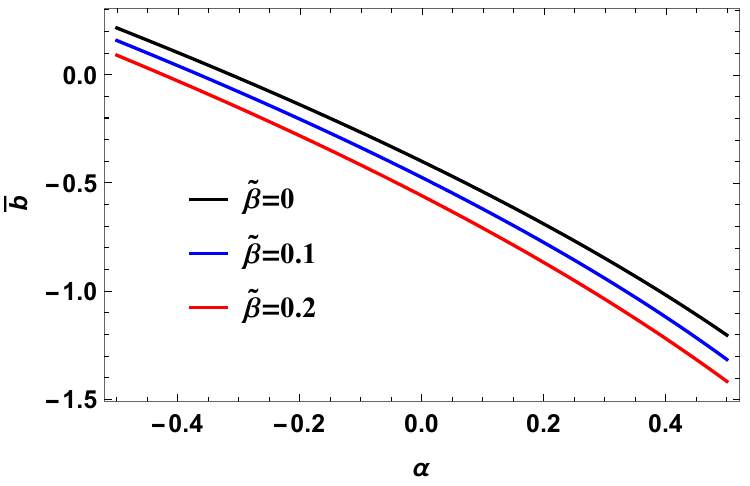}&
\includegraphics[width=0.4\columnwidth]{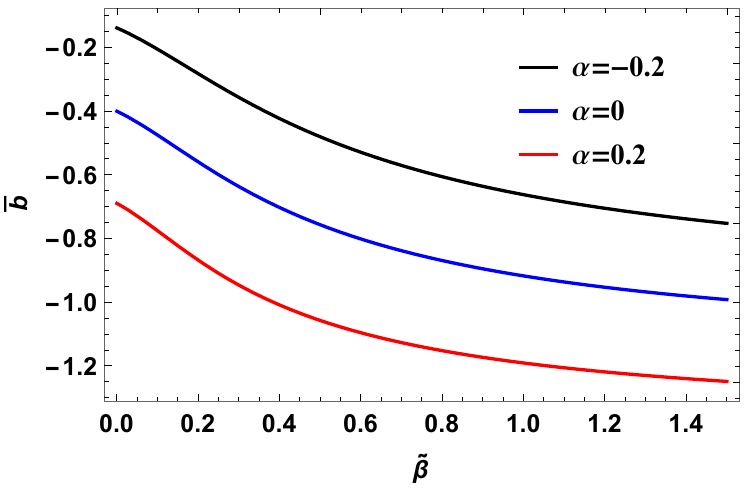}
\end{tabular}
\caption{Variation of lensing coefficient $\bb$ with $\a$ for different values of $\tb$ (left panel) and with $\tb$ for different values of $\a$ (right panel). }\label{bbar}
\end{center}
\end{figure}
Table (\r{abv}) accentuates the impact of parameters $\a$ and $\b$ on the lensing coefficients $\ab$ and $\bb$.
\begin{table}[H]
\begin{centering}
\begin{tabular*}{\textwidth}{@{\extracolsep{\fill}\quad}cccccc}
\hline\hline
{$\alpha$ }& {$\tb$} & {$\ab$} & {$\bb$} & {$\delta \ab$} & {$\delta \bb$}
\\ \hline
\hline
\multirow{5}{*}{-0.2}&$0$ & $1.09545$ & $-0.13858$ & $0.0954451$ & $0.26165$ \\
&$0.1$ & $1.0428$ & $-0.20527$ & $0.0427984$ & $0.19496$ \\
&$0.2$ & $0.989606$ & $-0.282183$ & $-0.0103943$ & $0.118047$ \\
&$0.3$ & $0.943469$ & $-0.356605$ & $-0.056531$ & $0.0436246$ \\
&$0.4$ & $0.905577$ & $-0.423045$ & $-0.0944233$ & $-0.0228149$ \\
\hline
\multirow{5}{*}{0}&$0$ & $1.$ & $-0.40023$ & $0.$ & $0.$ \\
&$0.1$ & $0.941914$ & $-0.474892$ & $-0.0580861$ & $-0.0746617$ \\
&$0.2$ & $0.885626$ & $-0.559169$ & $-0.114374$ & $-0.158939$ \\
&$0.3$ & $0.839648$ & $-0.636652$ & $-0.160352$ & $-0.236422$ \\
&$0.4$ & $0.803817$ & $-0.702368$ & $-0.196183$ & $-0.302138$ \\
\hline
\multirow{5}{*}{0.2}&$0$ & $0.894427$ & $-0.689643$ & $-0.105573$ & $-0.289413$ \\
&$0.1$ & $0.829211$ & $-0.775265$ & $-0.170789$ & $-0.375034$ \\
&$0.2$ & $0.770339$ & $-0.86766$ & $-0.229661$ & $-0.46743$ \\
&$0.3$ & $0.726218$ & $-0.946201$ & $-0.273782$ & $-0.545971$ \\
&$0.4$ & $0.694056$ & $-1.0084$ & $-0.305944$ & $-0.608174$ \\
\hline\hline
\end{tabular*}
\end{centering}
\caption{Values of lensing coefficients for different values of $\a$ and $\b$. \label{abv}}
\end{table}
In the above table, $\delta \ab$ and $\delta \bb$ are deviations of $\ab$ and $\bb$ from the \s case, respectively. For positive values of $\a$, lensing coefficients are always less than those for the \s BH. However, for negative values of $\a$, we can have $\ab$ and $\bb$ greater than those for the \s BH for small values of the PFDM parameter $\b$. The behavior of the angle of deflection as a function of the impact parameter is depicted in Fig. (\r{def}). The adverse impact of the KR field and PFDM on lensing is significant. This demonstrates a considerably weaker lensing effect by a KRPFDM BH than a \s BH. Thus, gravitational lensing can be utilized to distinguish between the two BHs.
\begin{figure}[H]
\begin{center}
\begin{tabular}{cc}
\includegraphics[width=0.4\columnwidth]{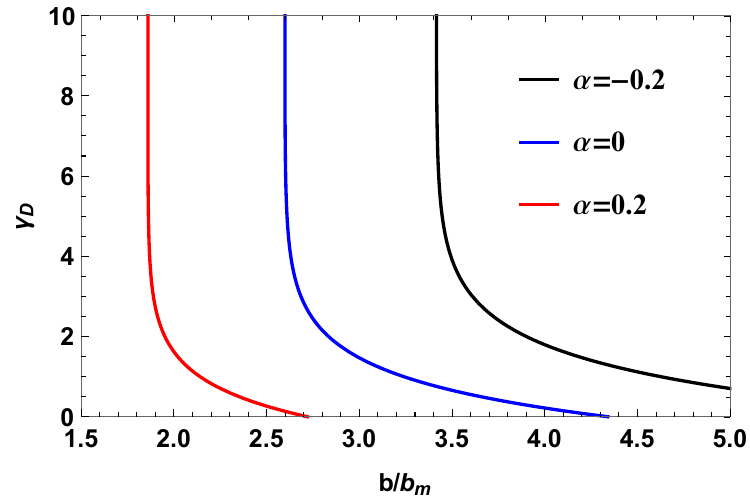}&
\includegraphics[width=0.4\columnwidth]{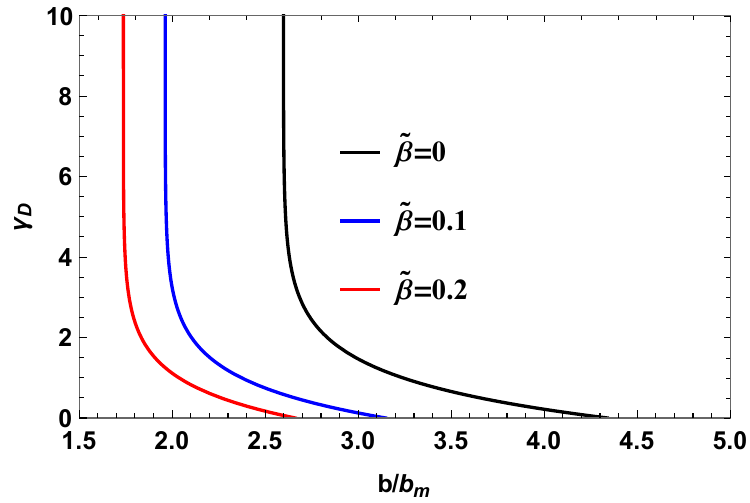}
\end{tabular}
\caption{Variation of deflection angle $\gamma_D$ as a function of $b/b_m$ for different values of $\a$ with $\tb=0$ (left panel) and for different values of $\tb$ with $\a=0$ (right panel). }\label{def}
\end{center}
\end{figure}
\section{observables in strong gravitational lensing}
This section explores lensing observables following articles \c{BOZZA, BOZZA1}. Here, we utilize the lens equation that connects the positions of the source S, the BH (lens L), and the observer O given by \c{BOZZA1}:
\begin{equation}
\eta=\theta-\frac{D_{LS}}{D_{OS}} \Delta \gamma_n,%
\label{lens}
\end{equation}
where $D_{LS}$ is the source to BH distance, $D_{OS}$ is the observer to source distance, $\Delta \gamma_n=\gamma(\theta)-2n\pi$ is the offset of the deflection angle, and $n$ is an integer that corresponds to winding number of loops around BH. Following \c{BOZZA}, the angular position occupied by the $n^{th}$ relativistic image is
\begin{equation}
\theta_n = \theta_n^0 +\frac{ b_m e_n\left(\eta-\theta_n^0\right)
D_{OS}}{\overline{a} D_{LS}D_{OL}},
\label{images}
\end{equation}
where $\theta_n^0$ is the solution of the equation $\gamma(\theta)=2n\pi$ with explicit expression given by
\begin{eqnarray}
\theta_n^0=\frac{b_m}{D_{OL}} \left(1+e_n \right) \label{theta0}\quad \text{where} \quad e_n=e^{\frac{\overline{b}-2n\pi}{\overline{a}}}.\label{en}
\end{eqnarray}
When $\theta_n^0$ equates $\eta$, the image and the source coincide, and hence the position of the $n^{th}$ image acquires no correction. Eq. (\r{images}) denotes images on the same side of the source. To obtain an image on the opposite side of the source, we can replace $\eta$ with $-\eta$. The magnification of the image encodes valuable information. It is defined by \c{BOZZA}
\begin{eqnarray}
\mu_n &=& \left(\frac{\eta}{\theta} \;
\;\frac{d\eta}{d\theta} \Bigg|_{\theta_n ^0}\right)^{-1}\\
&=&e_n \frac{ b_m^2\left(1+e_n \right)
D_{OS}}{\overline{a} \eta D_{OL}^2 D_{LS}}.\label{magnification}
\end{eqnarray}
The magnification of images exponentially decreases with the winding number $n$. As such, images with higher $n$ become faint unless $\eta$ tends to zero, implying a perfect alignment between the source and the lens. The outermost image at $\theta_1$ is considered to be resolved as a single image, whereas all other inner images are clustered at $\theta_\infty$. Following are the three observables that are being evaluated here:
\begin{eqnarray}
&&\theta_\infty = \frac{b_m}{D_{OL}},\quad s= \theta _1-\theta _\infty = \theta_\infty \;\; e^{\frac{\bar{b}-2\pi}{\bar{a}}},\quad \text{and \quad }r_{\text{mag}}= 2.5 \log(r) = \frac{5\pi}{\bar{a}\ln 10}\label{s}\\\nn
\text{where}\\
&&r = \frac{\mu_1}{\sum{_{n=2}^\infty}\mu_n } = e^{\frac{2 \pi}{\bar{a}}}.\label{obs}
\end{eqnarray}
Here, the first image and the rest have an angular separation of $s$, $r_{mag}$ describes the relative magnification of the first image in comparison with the rest, and $r$ describes the ratio of the flux from the outermost image to the flux from the rest. The relative magnification, $r_{mag}$, depends neither on the BH mass nor on the BH's distance from the observer. With $\theta_\infty$, $s$, and $r_{mag}$ at our disposal from astronomical observations, we can evaluate values of $\ab$, $\bb$, and $b_m$ from Eq. ({\r{obs}) that will shed light on the nature of the KRPFDM BH.
\begin{figure}[H]
\begin{center}
\begin{tabular}{cc}
\includegraphics[width=0.4\columnwidth]{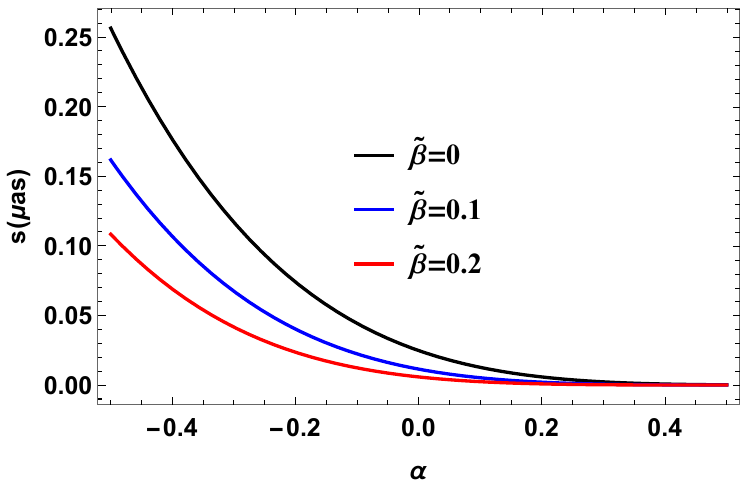}&
\includegraphics[width=0.4\columnwidth]{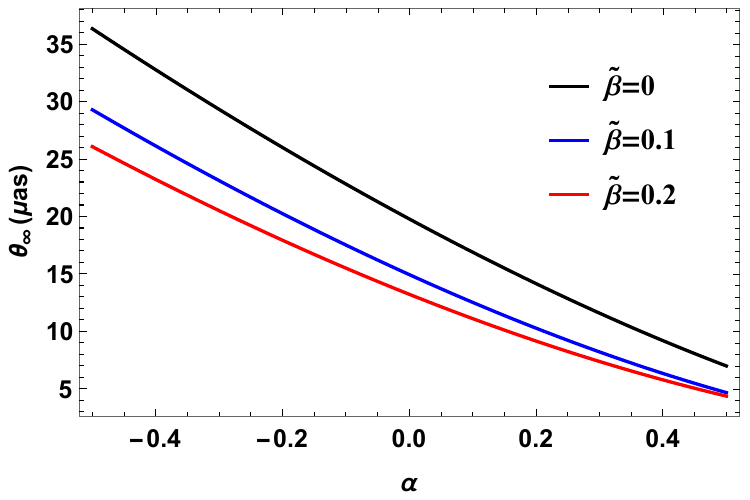}
\end{tabular}
\caption{Variation of observables angular separation $s$ (left panel) and angular position $\theta_\infty$ (right panel) as a function of $\a$ for different values of $\tb$ by modelling the supermassive BH $M87^*$ as a KRPFDM BH. }\label{sm871}
\end{center}
\end{figure}
\begin{figure}[H]
\begin{center}
\begin{tabular}{cc}
\includegraphics[width=0.4\columnwidth]{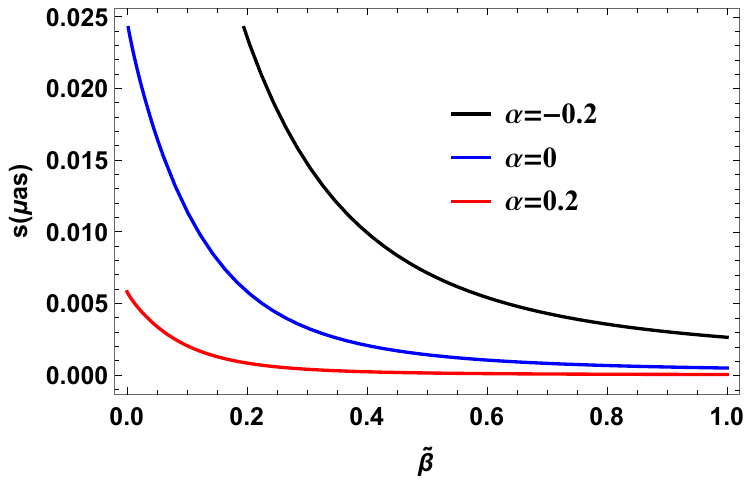}&
\includegraphics[width=0.4\columnwidth]{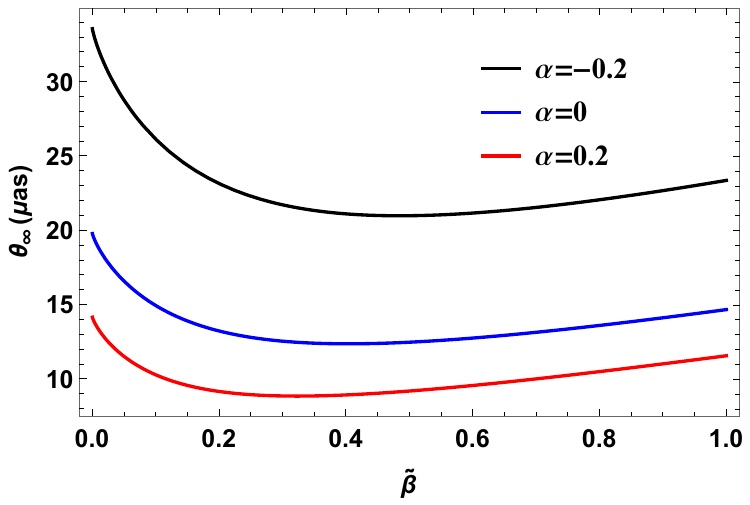}
\end{tabular}
\caption{Variation of observables angular separation $s$ (left panel) and angular position $\theta_\infty$ (right panel) as a function of $\tb$ for different values of $\a$ by modelling the supermassive BH $M87^*$ as a KRPFDM BH. }\label{sm872}
\end{center}
\end{figure}
\begin{figure}[H]
\begin{center}
\begin{tabular}{cc}
\includegraphics[width=0.4\columnwidth]{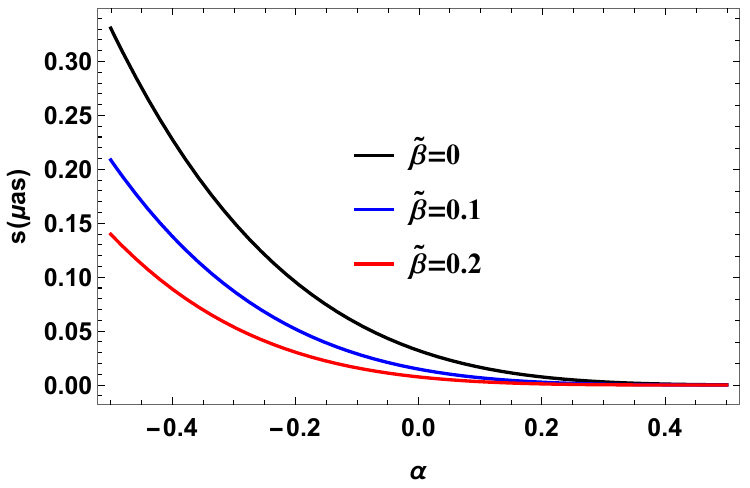}&
\includegraphics[width=0.4\columnwidth]{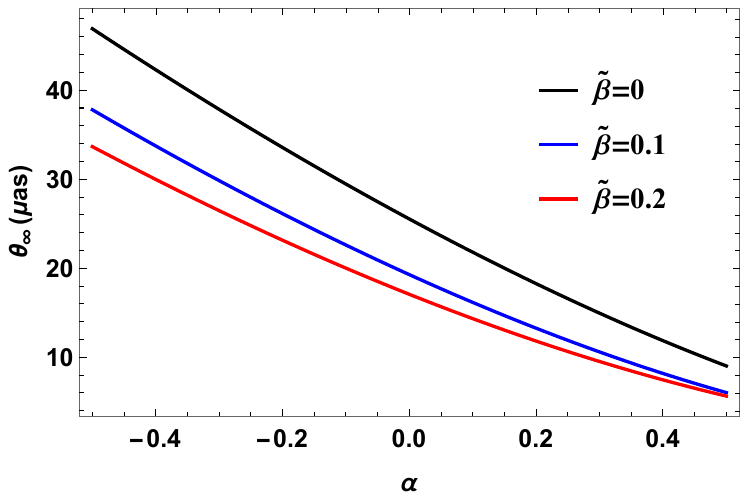}
\end{tabular}
\caption{Variation of observables angular separation $s$ (left panel) and angular position $\theta_\infty$ (right panel) as a function of $\a$ for different values of $\tb$ by modelling the supermassive BH $SgrA^*$ as a KRPFDM BH. }\label{ssgra1}
\end{center}
\end{figure}
\begin{figure}[H]
\begin{center}
\begin{tabular}{cc}
\includegraphics[width=0.4\columnwidth]{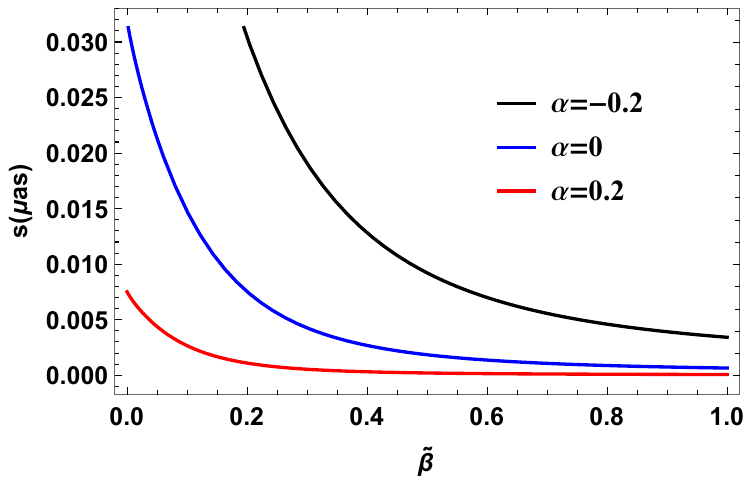}&
\includegraphics[width=0.4\columnwidth]{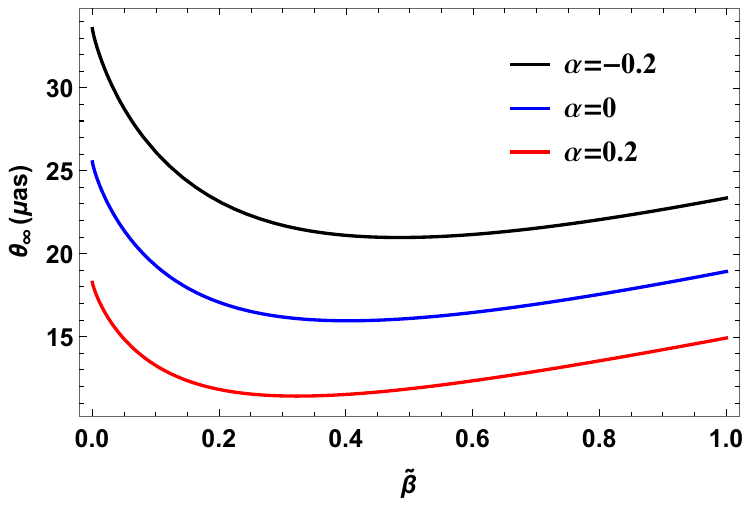}
\end{tabular}
\caption{Variation of observables angular separation $s$ (left panel) and angular position $\theta_\infty$ (right panel) as a function of $\tb$ for different values of $\a$ by modelling the supermassive BH $SgrA^*$ as a KRPFDM BH. }\label{ssgra2}
\end{center}
\end{figure}
Fig. (\r{sm871}), (\r{sm872}), (\r{ssgra1}), and (\r{ssgra2}) demostrate qualitative impact of parameters $\a$ and $\tb$ on strong obsevables $s$ and $\theta_\infty$ where we have modelled SMBHs $M87^*$ and $SgrA^*$ as KRPFDM BHs. These two observables diminish with an increase in $\a$. Critical behaviour of the angular position $\theta_\infty$ is conspicuous from Fig. (\r{sm872}) and (\r{ssgra2}). Here, we see that the angular position reaches its minimum value of $20.9772 \mu as$ ($SgrA^*$) and $16.2527 \mu as$ ($M87^*$) at $\tb=0.484095$ when $\a=-0.2$, attains minimum value of $15.9597 \mu as$ ($SgrA^*$) and $12.3639 \mu as$ ($M87^*$) at $\tb=0.403412$ when $\a=0$ and gets minimum value of $11.4185 \mu as$ ($SgrA^*$) and $8.84687 \mu as$ ($M87^*$) at $\tb=0.32273$ when $\a-0.2$. The behavior of the relative magnification as a function of $\a$ and $\tb$ are illustrated in Fig. (\r{rmag}). It displays incremental behavior of this lesnsing observable with increasing either $\a$ or $\b$.
\begin{figure}[H]
\begin{center}
\begin{tabular}{cc}
\includegraphics[width=0.4\columnwidth]{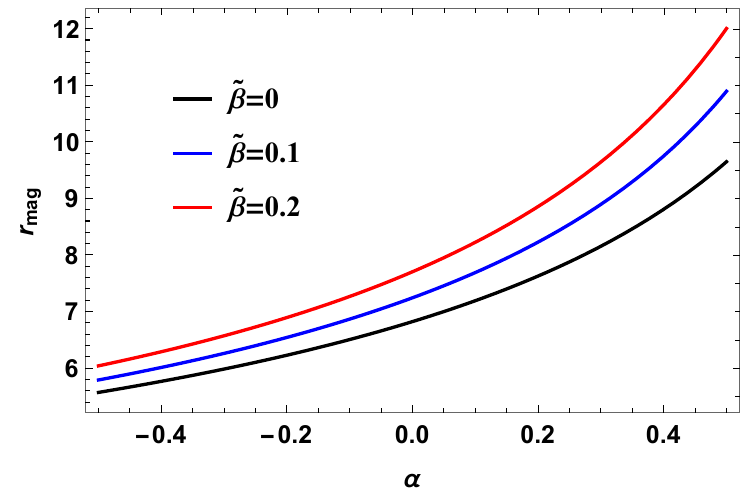}&
\includegraphics[width=0.4\columnwidth]{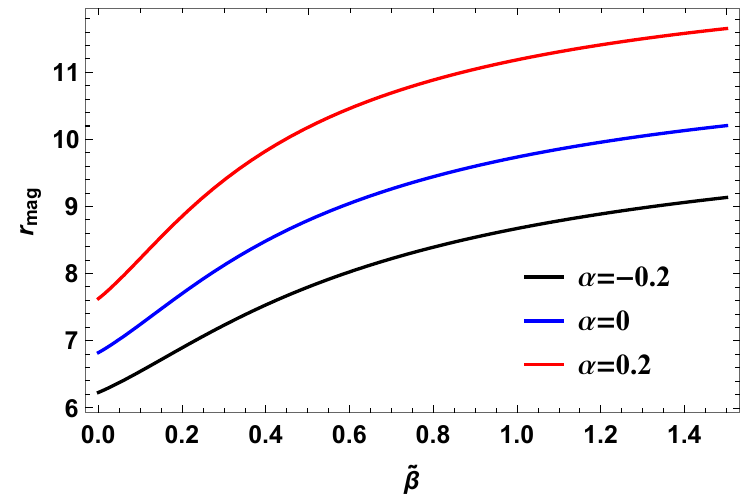}
\end{tabular}
\caption{Variation of relative magnification $r_{mag}$ as a function of $\a$ for different values of $\tb$ (left panel) and as a function of $\tb$ for different values of $\a$ (right panel)}\label{rmag}
\end{center}
\end{figure}
Having explored the qualitative aspect, we now demonstrate the quantitative extent of influence parameters $\a$ and $\tb$ have on strong observables.
\begin{table}[H]
\begin{centering}
\setlength{\tabcolsep}{0pt}
\begin{tabular*}{\textwidth}{@{\extracolsep{\fill}\quad}lccccccc }
\hline\hline
\multicolumn{2}{c}{}&
\multicolumn{2}{c}{Sgr A*}&
\multicolumn{2}{c}{M87*}& \\
{$\alpha$ }& {$\tb$} & {$\theta_\infty $ ($\mu$as)} & {$s$ ($\mu$as) } & {$\theta_\infty $ ($\mu$as)} & {$s$ ($\mu$as) } & {$r_{mag} $ }
\\ \hline
\hline
\multirow{5}{*}{-0.2}&$0$ & $33.5598$ & $0.0954726$ & $26.0015$ & $0.0739705$ & $6.2275$  \\
&$0.1$ & $26.1287$ & $0.0518644$ & $20.244$ & $0.0401836$ & $6.5419$  \\
&$0.2$ & $23.1439$ & $0.0304218$ & $17.9315$ & $0.0235702$ & $6.89353$  \\
&$0.3$ & $21.7131$ & $0.0190684$ & $16.8229$ & $0.0147739$ & $7.23064$ \\
&$0.4$ & $21.1056$ & $0.0128303$ & $16.3522$ & $0.0099407$ & $7.53319$ \\
\hline
\multirow{5}{*}{0}&$0$ & $25.5298$ & $0.0319504$ & $19.78$ & $0.0247546$ & $6.82188$ \\
&$0.1$ & $19.2797$ & $0.0147608$ & $14.9376$ & $0.0114364$ & $7.24257$ \\
&$0.2$ & $17.0704$ & $0.00753145$ & $13.2258$ & $0.00583523$ & $7.7029$ \\
&$0.3$ & $16.1856$ & $0.00426534$ & $12.5403$ & $0.00330471$ & $8.12469$ \\
&$0.4$ & $15.9581$ & $0.00268383$ & $12.364$ & $0.00207938$ & $8.48686$ \\
\hline
\multirow{5}{*}{0.2}&$0$ & $18.2676$ & $0.00751594$ & $14.1534$ & $0.00582321$ & $7.6271$  \\
&$0.1$ & $13.2634$ & $0.00266582$ & $10.2762$ & $0.00206543$ & $8.22696$  \\
&$0.2$ & $11.8191$ & $0.00109938$ & $9.15722$ & $0.000851781$ & $8.85569$ \\
&$0.3$ & $11.4291$ & $0.00054285$ & $8.85508$ & $0.00042059$ & $9.39371$  \\
&$0.4$ & $11.5178$ & $0.000315338$ & $8.92378$ & $0.000244318$ & $9.829$ \\
\hline\hline
\end{tabular*}
\end{centering}
\caption{Strong-lensing observables for supermassive black holes $Sgr A*$
and $M87^*$. \label{table2}}
\end{table}
For $\a \geq 0$, $s$ and $\theta_\infty$ are always less than those for a \s BH. However, for negative values of $\a$, the angular separation between the outermost image and the inner-packed images $s$ and the angular position of the inner-packed images $\theta_\infty$ can be greater than \s values depending on the PFDM parameter $\tb$. Increasing either parameter increases the flux from the outermost image compared to the flux from the rest of the images. This concludes our study of strong gravitational lensing in the background of KRPFDM BHs, which evinces the significant impact of LSB and PFDM on lensing observables.
\section{parameter estimation using shadow observable}
This section utilizes experimental observations related to shadows of the SMBHs $M87^*$ (by EHT) and $SgrA^*$ (by the Keck and VLTI observatories) to test the feasibility of our model and put constraints on model parameters \cite{M871, M872, M873, keck, vlti1, vlti2, vlti3}. Here, we employ bounds on the parameter $\delta$, deviation from Schwarzschild, to our end. It is defined by \c{del}
\be
\delta=\f{R_s}{3\sqrt{3}M}-1,
\ee
where $R_s$ is the shadow radius connected to the critical impact parameter $b_m$ through the following relation
\be
R_s \lt(\\a, \b \rt)=2M b_m \lt(\a, \f{\b}{2M}\rt).
\ee
Its value in the limit $\a \rightarrow 0$ and $\b \rightarrow 0$ provides $3\sqrt{3}$, the shadow radius for a \s BH, and hence the parameter $\delta$ becomes zero. An overview of $\delta$'s dependence on our model parameters $\a$ and $\b$ is demonstrated in Fig. (\r{del}). Even though the deviation parameter decreases linearly with $\a$, it exhibits critical behavior with respect to the variation of PFDM parameter $\b$. Interestingly, apart from $(\a,\, \b)=(0,\, 0)$, we have other combinations of $(\a,\, \b)$ that also generate $\delta=0$, i.e., the shadow radius for a KRPFDM BH in such cases equates the shadow radius of a \s BH. Such combinations only occur for negative values of $\a$. Some such combinations of $(\a,\, \b)$ are $(-0.15,\, 0.147027M)$, $(-0.1,\, 0.0824817M)$, and $(-0.05,\, 0.033025M)$. In such circumstances, the effect of non-zero VEV of the KR field gets nullified by the PFDM, and a KRPFDM BH produces a shadow of the same size as that of a \s BH in a vacuum.
\begin{figure}[H]
\begin{center}
\begin{tabular}{cc}
\includegraphics[width=0.4\columnwidth]{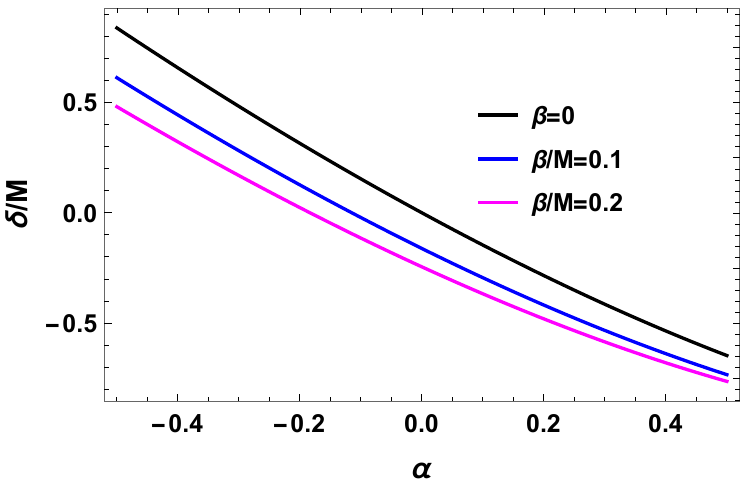}&
\includegraphics[width=0.4\columnwidth]{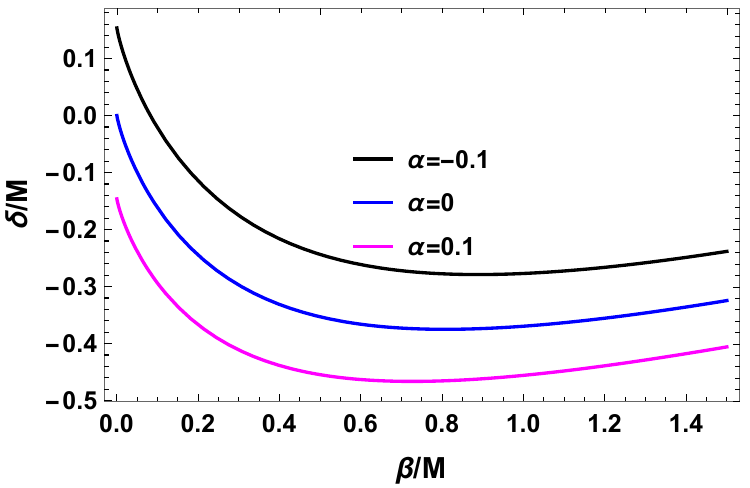}
\end{tabular}
\caption{Variation of deviation from \s $\delta$ as a function of $\a$ for different values of $\b$ (left panel) and as a function of $\b$ for different values of $\a$ (right panel)}\label{del}
\end{center}
\end{figure}
Bounds on $\delta$ reported by the EHT, Keck, and VLTI observatories are given in Table (\ref{bounds}).
\begin{center}
\begin{tabular}{|l|c|c|c|r|}
\hline
BH & Observatory & $\delta$ & 1$\sigma$ bounds & 2$\sigma$ bounds\\
\hline
\hline
$M87^*$ & EHT & $-0.01^{+0.17}_{-0.17}$ & $4.26\le \frac{R_s}{M}\le 6.03$ &  $3.38\le \frac{R_s}{M}\le 6.91$\\
\hline
\hline
\multirow{2}{*}{$Sgr A^*$}&{VLTI} & $-0.08^{+0.09}_{-0.09}$ &{$4.31\le \frac{R_s}{M}\le 5.25$} &{$3.85\le \frac{R_s}{M}\le 5.72$}\\[3mm]
& Keck & $-0.04^{+0.09}_{-0.10}$ & {$4.47\le \frac{R_s}{M}\le 5.46$} & {$3.95\le \frac{R_s}{M}\le 5.92$}\\[1mm]
\hline
\end{tabular}
\captionof{table}{Bounds on $\delta$ from different observatories.} \label{bounds}
\end{center}
We intend to find combined bounds for our model parameters that satisfy the reported constraints on $\delta$ within a $1 \sigma$ confidence level (CL). Modeling SMBHs $M87^*$ and $SgrA^*$ as KRPFDM BHs, we present in Fig. (\r{fm87}, (\r{fsgra1}), and (\r{fsgra2}) parameter space that produces theoretical results consistent with experimental values within $1\sigma$ CL.
\begin{figure}[H]
\begin{center}
\begin{tabular}{cc}
\includegraphics[width=0.4\columnwidth]{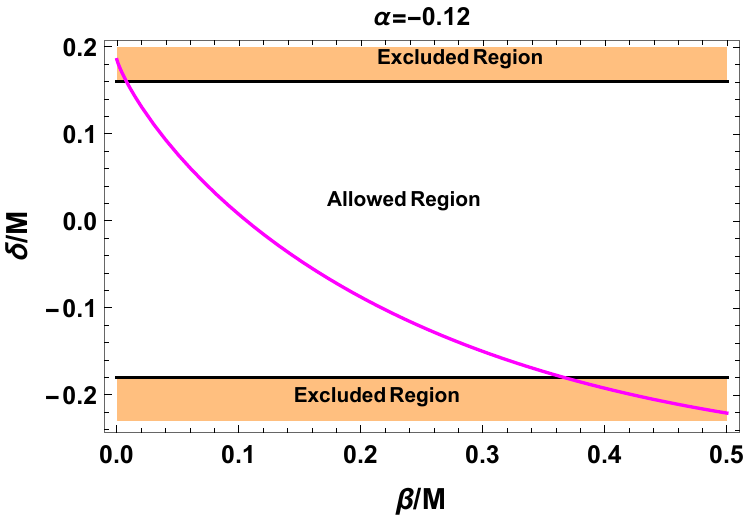}&
\includegraphics[width=0.4\columnwidth]{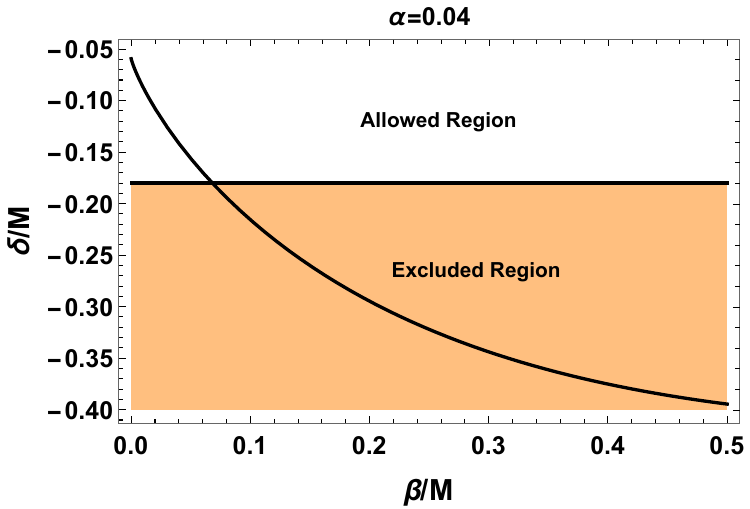}
\end{tabular}
\caption{Variation of deviation from \s $\delta$ as a function of $\b$ for $\a=-0.12$ (left panel) and for $\a=0.04$ (right panel). Our metric is concordant with the EHT results within $1\sigma$ CL for white regions.}\label{fm87}
\end{center}
\end{figure}
\begin{figure}[H]
\begin{center}
\begin{tabular}{cc}
\includegraphics[width=0.4\columnwidth]{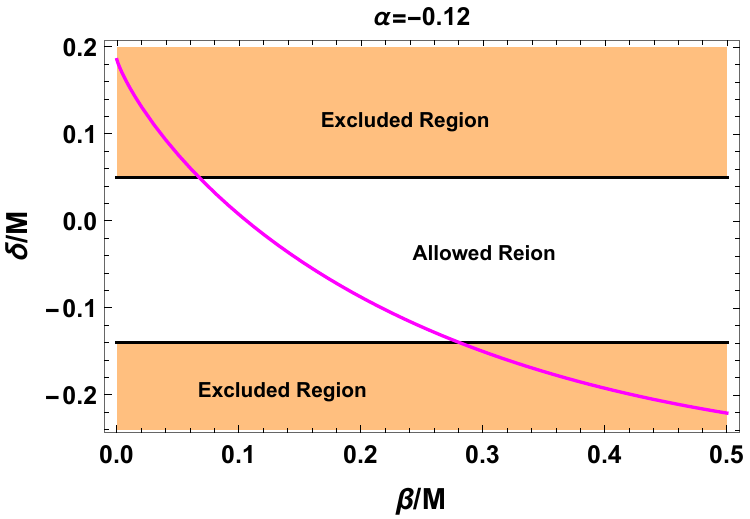}&
\includegraphics[width=0.4\columnwidth]{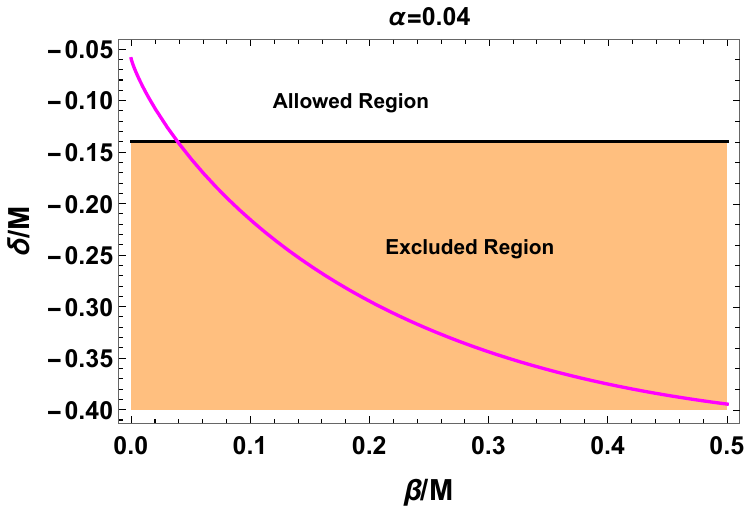}
\end{tabular}
\caption{Variation of deviation from \s $\delta$ as a function of $\b$ for $\a=-0.12$ (left panel) and for $\a=0.04$ (right panel). Our metric is concordant with the Keck results within $1\sigma$ CL for white regions.}\label{fsgra1}
\end{center}
\end{figure}
\begin{figure}[H]
\begin{center}
\begin{tabular}{cc}
\includegraphics[width=0.4\columnwidth]{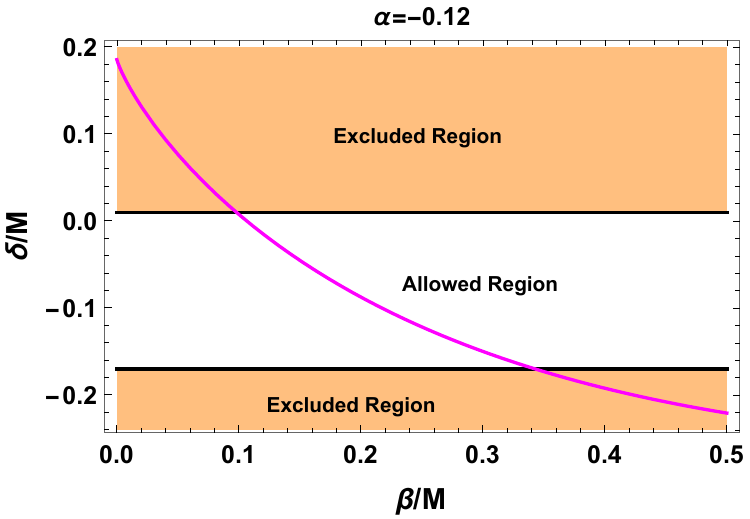}&
\includegraphics[width=0.4\columnwidth]{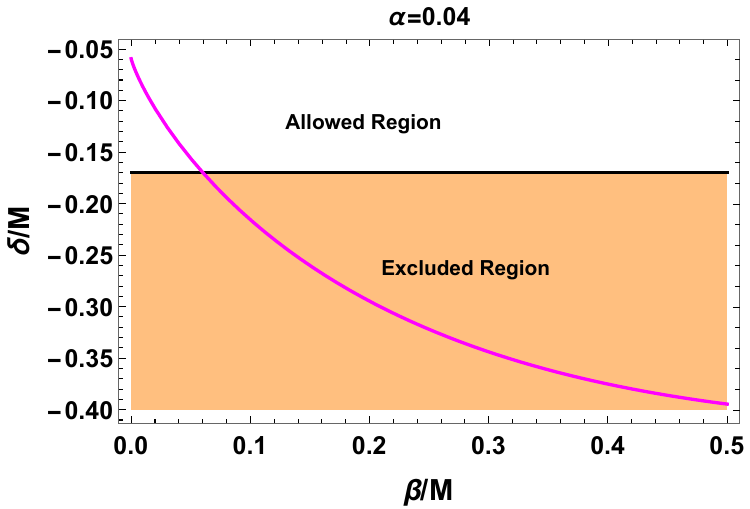}
\end{tabular}
\caption{Variation of deviation from \s $\delta$ as a function of $\b$ for $\a=-0.12$ (left panel) and for $\a=0.04$ (right panel). Our metric is concordant with the VLTI results within $1\sigma$ CL for white regions.}\label{fsgra2}
\end{center}
\end{figure}
Table (\r{ob}) provides constraints on parameters $\a$ and $\b$ that we have extracted by comparing our theoretical results with experimental findings. As expected, the EHT results provide the largest range of possible values of our model parameters. Our study puts stringent bounds on parameters $\a$ and $\b$ and makes our model a feasible candidate as an SMBH.
\begin{table}[H]
\begin{centering}
\setlength{\tabcolsep}{0pt}
\begin{tabular*}{\textwidth}{@{\extracolsep{\fill}\quad}lccccccc }
\hline\hline
\multicolumn{1}{c}{}&
\multicolumn{1}{c}{}&
\multicolumn{2}{c}{Bounds on $\a$}&
\multicolumn{2}{c}{Bounds on $\b/M$}\\
{SMBH}& {Observatory} & {Lower bound} & {Upper bound } & {Lower Bound} & {Upper bound}
\\
\\
\hline\\
\multirow{1}{*}{$M87^*$}&EHT & $-0.104008$ & $0.057175$ & $0$ & $0.0500495$ \\
\\
\hline\\
\multirow{2}{*}{$SgrA^*$}& Keck & $-0.0330621$ & $0.042171$ & $0$ & $0.0370688$\\
\\
&VLTI & $-0.00665618$ & $0.051$ & $0$ & $0.0489939$  \\
\\
\hline\hline
\end{tabular*}
\end{centering}
\caption{Bounds on $\a$ and $\b$ extracted from experimental results.}\label{ob}
\end{table}
\section{conclusions}
This article is devoted to finding a static and spherically symmetric metric that combines the effects of LSB and the presence of PFDM. The Lorentz symmetry is spontaneously broken in our case due to the non-zero VEV of the KR field. The KR field is considered to be frozen to its VEV that spontaneously breaks particle Lorentz symmetry. Modified field equations were solved, and the exact solution was obtained that exhibited new properties owing to the combined effects of LSB and PFDM. In order to analyze the nature of singularities, expressions of scalar invariants, Ricci Scalar, Ricci squared, and Kretschmann Scalar were obtained. Our metric was found to be non-Ricci flat owing to non-zero components of the Ricci tensor.\\
We then moved on to explore the combined effect of LSB and PFDM on strong gravitational lensing and its observables. To this end, We followed the prescription detailed in \cite{BOZZA, BOZZA1, BOZZA2}. Our study revealed that the event horizon, photon radius, and critical impact parameter decreased linearly with the parameter $\a$. In contrast, they exhibited critical behavior with respect to their variations against $\b$. The lensing coefficients, as well as the deflection angle, were found to be diminishing with increasing either $\a$ or $\b$. Strong lensing observables, the angular position $\theta_\infty$, angular separation $s$, and relative magnification $r_{mag}$, were also explored to gauge the impact of parameters $\a$ and $\b$. For $\a \geq 0$, these observables were always found to be less than those for a \s BH for any positive value of $\b$. However, for negative values of $\a$, we may have these observables greater than, equal to, or less than the \s values depending on the value of $\b$. \\
Finally, we employed bounds on the deviation from Schwarzschild, $\delta$, obtained by the EHT for the SMBH $M87^*$ and by the Keck and VLTI observatories for the SMBH $SgrA^*$ to extract possible values of our model parameters $\a$ and $\b$ that would make our model concomitant with experimental observations. Bounds obtained from our analysis are displayed in Table (\r{ob}). Our study revealed that, for a finite parameter space $(\a, \b)$, our model is consistent with observations, making it a feasible candidate for SMBHs. However, with the current observations, it is not possible to distinguish a KRPFDM BH and a \s BH. With finer results expected from ngEHT, we will possibly be able to extract exact constraints.

\end{document}